\documentclass[10pt,a4paper]{article}

\usepackage[margin=2.5cm]{geometry}
\usepackage{times}
\usepackage{graphicx}
\usepackage{amsmath,amssymb}
\usepackage{bbm}
\usepackage{authblk}
\usepackage[numbers,sort&compress]{natbib}
\usepackage{hyperref}
\title{\bfseries Conformal and Noncommutative Gravity from Gauge Symmetry Breaking and SO(2,16) Unification}

\author[1]{Stelios Stefas}
\author[1,2,3,4]{George Zoupanos}
\affil[1]{\small Physics Department, National Technical University of Athens, Zografou Campus, 157 80, Zografou, Greece}
\affil[2]{\small Max-Planck-Institut f\"ur Physik, Boltzmannstr.\ 8, 85748 Garching/Munich, Germany}
\affil[3]{\small Universit\"at Hamburg, Luruper Chaussee 149, 22761 Hamburg, Germany}
\affil[4]{\small Deutsches Elektronen-Synchrotron DESY, Notkestra{\ss}e 85, 22607 Hamburg, Germany}
\affil[ ]{\small E-mails: \href{mailto:dstefas@mail.ntua.gr}{dstefas@mail.ntua.gr}, \href{mailto:george.zoupanos@cern.ch}{george.zoupanos@cern.ch}}

\date{} 

\begin{document}

\maketitle

\begin{abstract}
  Motivated by the old idea of treating gravity as a gauge theory, we review the gauge-theoretic construction of Conformal Gravity and of Noncommutative (Fuzzy) Gravity. We then outline how these gauge formulations of gravity can be unified with internal interactions described by an $SO(10)$ Grand Unified Theory, based on the critical observation that the dimension of the tangent-space symmetry group of a curved manifold need not coincide with the spacetime dimension.
\end{abstract}

\section{Introduction}

For more than a century, unification has been a recurring theme in theoretical physics, revisited with different tools and in different guises. A classic starting point is the Kaluza Klein proposal to unify gravity and electromagnetism, the two established interactions at the time, by extending spacetime to five dimensions~\cite{Kaluza:1921,Klein:1926}, i.e.\ by using extra dimensions as a unification device. Higher-dimensional ideas gained new momentum once it was realised that non-Abelian gauge theories can emerge naturally in such settings~\cite{Kerner:1968, CHO1987358, Cho:1975sf}, opening the possibility of describing gravity together with the Standard Model (SM) in a single geometric framework. More concretely, if spacetime is taken to be a direct product $M_D=M_4\times B$, where $B$ is a compact Riemannian manifold with a non-Abelian isometry group $S$, then dimensional reduction to four dimensions yields gravity coupled to a Yang-Mills theory with gauge group $S$, accompanied by scalar fields. The attractiveness of this construction lies in its geometrical interpretation of gauge symmetries and in the hope of unifying gravity with the remaining interactions. The minimal setup, however, runs into serious difficulties: it typically lacks a viable classical ground state compatible with the simple direct-product ansatz and, even more importantly for low-energy physics, it fails to produce chiral fermions in four dimensions after reduction~\cite{Witten:1983}. A significant improvement is obtained by introducing Yang-Mills fields already at the higher-dimensional level, sacrificing the purely geometric unification while keeping the extra-dimensional setting. This led to the Coset Space Dimensional Reduction (CSDR) programme~\cite{forgacs, MANTON1981502, Kubyshin:1989vd, KAPETANAKIS19924}, introduced by Forg\'acs and Manton (F-M), which can naturally yield chiral fermions in four dimensions. In a closely related construction, Scherk and Schwarz (S-S) considered reduction on group manifolds~\cite{SCHERK197961}; although chirality is not accommodated there, the idea has played an important role in subsequent string model building. A clear lesson from these developments is that, in higher-dimensional Grand Unified Theory (GUT) frameworks that include both Yang-Mills fields and fermions~\cite{Georgi:1974sy, FRITZSCH1975193}, the emergence of chiral fermions in four dimensions requires the total spacetime dimension to be of the form $4n+2$~\cite{CHAPLINE1982461}.

Superstring Theories (ST) (see, e.g., Refs.~\cite{Green2012-ul, polchinski_1998, Lust:1989tj}), developed somewhat later with the same unification goal, dominated the study of higher-dimensional unification for decades. They provide a consistent higher-dimensional framework, with the heterotic string theory~\cite{GROSS1985253}, formulated in ten dimensions, being a particularly attractive example. It naturally accommodates GUT gauge groups such as $E_8 \times E_8$, whose dimensional reduction can, in principle, reproduce the SM. At the same time, it should be emphasised that experimental support for higher-dimensional scenarios remains absent, despite recent and very promising developments within the CSDR framework~\cite{Manousselis_2004, Chatzistavrakidis:2009mh, Irges:2011de, Manolakos:2020cco, Patellis:2024dfl}.

Before the revival of higher-dimensional approaches, an additional important development had already been done directly in four dimensions. It provided a natural link between gravity and gauge theories, and thus pointed again towards unification. In particular, the SM is a gauge theory, and it has long been understood that gravity can also be formulated as a gauge theory~\cite{utiyama, kibble1961, Sciama, Umezawa, Matsumoto, macdowell, Ivanov:1980tw, Ivanov:1981wn, stellewest, Kibble:1985sn}. Interest in this viewpoint was strengthened by advances in supergravity~\cite{freedman_vanproeyen_2012, Ortín_2015}, which likewise rely on gauge principles, and more recently it has been extended to Noncommutative (NC) gravity~\cite{castellani, Chatzistavrakidis_2018, Manolakos_paper1, manolakosphd, Manolakos_paper2, Manolakos:2022universe, Manolakos:2023hif, roumelioti2407}.

Regarding gauge formulations of gravity, Weyl~\cite{weyl, weyl1929} took an initial step by relating electromagnetism to local phase transformations of the electron field and by introducing the vierbein formalism, which later became central in gauge-theoretic constructions of gravity. Utiyama~\cite{utiyama} subsequently argued that gravity may be treated as a gauge theory of the Lorentz group $SO(1,3)$, although the vierbein entered his formulation in a somewhat ad hoc manner. This point was addressed by Kibble~\cite{kibble1961} and Sciama~\cite{Sciama}, who proposed gauging the full Poincar\'e group. Later work by Stelle and West~\cite{stellewest, Kibble:1985sn} yielded more refined models based on the de Sitter $SO(1,4)$ or anti-de Sitter $SO(2,3)$ groups, which are related to the Poincar\'e group by having the same number of generators. Then, employing spontaneous symmetry breaking (SSB) Lorentz invariance is recovered. The conformal group $SO(2,4)$ also played a key role in the development of conformal (CG) and Weyl (WG) gravities~\cite{KAKU1977304, Roumelioti:2024lvn}, fuzzy gravity (FG)~\cite{Chatzistavrakidis_2018, Manolakos_paper1, manolakosphd, Manolakos_paper2, Manolakos:2022universe, Manolakos:2023hif, roumelioti2407}, and their supersymmetric extensions in $N=1$ supergravity~\cite{KAKU1977304, freedman_vanproeyen_2012}.

A more direct and ambitious proposal, aiming at a unified description of all interactions within a gauge-theoretic framework, has recently been revived~\cite{Nesti_2008, Nesti_2010, Chamseddine2010, Chamseddine2016, Krasnov:2017epi, Konitopoulos:2023wst, Manolakos:2023hif, noncomtomos, roumelioti2407, Roumelioti:2024lvn, Patellis-Z-24, Roumelioti:2025cco, Roumelioti:2025cxi, Patellis:2025qbl}, motivated by earlier attempts~\cite{Weinberg:1984ke, Percacci:1984ai, Percacci_1991}. The key observation is that the dimension of the tangent group of a curved spacetime does not necessarily coincide with the dimension of the manifold itself. This opens the possibility of using higher-dimensional tangent groups over four-dimensional spacetime, thereby enabling a unified gauge-theoretic treatment of gravity together with internal interactions. In this way, techniques originally developed for higher-dimensional theories, such as CSDR~\cite{CHAPLINE1982461, forgacs, MANTON1981502, Kubyshin:1989vd, KAPETANAKIS19924, LUST1985309, Manousselis_2004, Chatzistavrakidis:2009mh, Irges:2011de, Manolakos:2020cco, Patellis:2024dfl}, can be adapted to a four-dimensional setting. Nevertheless, difficulties, including the simultaneous implementation of Weyl and Majorana conditions in order to obtain realistic chiral spectra, appear in this formulation~\cite{CHAPLINE1982461, KAPETANAKIS19924}. More recently, unified gauge frameworks have been constructed that bring together conformal gravity and internal interactions, and have been extended further to unify noncommutative (fuzzy) gravity in a similar manner~\cite{Manolakos:2023hif,roumelioti2407, Konitopoulos:2023wst, Patellis-Z-24, Roumelioti:2024lvn, Roumelioti:2025cco, Roumelioti:2025cxi, Patellis:2025qbl}.

\section{Gauge-Theoretic Construction of Conformal Gravity}
\label{sec2}

Even though it is well known that Einstein gravity (EG) itself can already be expressed as a gauge theory of the Poincar\'e group~\cite{kibble1961}, a more elegant and symmetric starting point is provided by the ten-generator de Sitter $SO(1,4)$ or Anti-de Sitter $SO(2,3)$ groups, which match the Poincar\'e group in generator count and recover the Lorentz group $SO(1,3)$ through SSB driven by a suitable scalar field~\cite{stellewest, Kibble:1985sn, Roumelioti:2024lvn, manolakosphd}. All three of these groups sit inside the fifteen-generator conformal group $SO(2,4)$, and gauging this larger symmetry directly~\cite{Kaku:1978nz} produces conformal gravity (CG). Whereas the original formulation reduced CG down to EG or to Weyl's scale-invariant gravity by imposing algebraic constraints on the gauge fields by hand, a genuinely dynamical realisation of this breaking - through an action-level scalar field together with a Lagrange-multiplier constraint - was achieved only more recently~\cite{Roumelioti:2024lvn}.

Since $SO(2,4)$ is isomorphic to both $SU(4)$ and $SO(6)$, it is convenient to work in Euclidean signature from here on; in this language CG's gauge group admits two genuinely different symmetry-breaking paths down to $SO(1,3)$.

\paragraph{Route I: Breaking with Scalars in the Vector Representation}\hfill \break
The route that leads to EG proceeds by giving a vev to a scalar in the vector representation $\mathbf{6}$ of $SO(6)$, oriented along its $\langle \mathbf{1} \rangle$ singlet direction; this is possible because of the branching
\begin{equation}
  \label{SO6toSO5}
  \begin{aligned}
    SO(6) &\supset SO(5),\\
    \mathbf{6} &= \mathbf{1} + \mathbf{5}\, .
  \end{aligned}
\end{equation}
A second vev, this time for a scalar in the $\mathbf{5}$ of the surviving $SO(5)\sim SO(2,3)$, completes the reduction to the Lorentz group via
\begin{equation}
  \label{SO5toSU2SU2}
  \begin{aligned}
    SO(5) &\supset SU(2) \times SU(2),\\
    \mathbf{5} &= (\mathbf{1},\mathbf{1}) + (\mathbf{2},\mathbf{2})\, ,
  \end{aligned}
\end{equation}
since $SU(2)\times SU(2)$ here coincides with both $SO(4)$ and the Lorentz algebra $SO(1,3)$. The full chain $SO(2,4)\longrightarrow SO(1,3)$ thus requires exactly two scalar vevs in the $\mathbf{6}$, as described in~\cite{Roumelioti:2024lvn}.

\paragraph{Route II: One-Step Breaking with the Antisymmetric Representation}\hfill \break
The second route reaches $SO(1,3)$ in a single step, using instead a scalar in the second-rank antisymmetric representation $\mathbf{15}$ of $SO(6)\cong SO(2,4)$; depending on which vacuum is chosen within this representation, the outcome is either EG or Weyl gravity (WG), as explained below.

In four-dimensional language the fifteen conformal generators split into
\[
  \{M_{ab},\; P_a,\; K_a,\; D\},
\]
Lorentz transformations, translations, special conformal transformations, and dilatations, respectively, and the corresponding $SO(2,4)$ gauge connection decomposes as
\begin{equation}
  A_\mu
  = \frac{1}{2}\omega_\mu{}^{ab} M_{ab}
  + e_\mu{}^a P_a
  + b_\mu{}^a K_a
  + \tilde{a}_\mu D ,
\end{equation}
with $e_\mu{}^a$, $\omega_\mu{}^{ab}$, $b_\mu{}^a$ and $\tilde{a}_\mu$ respectively the vierbein, spin connection, special conformal gauge field, and dilatation gauge field. Its field strength follows as
\begin{equation}
  \label{fieldstrengthconformal}
  F_{\mu\nu}
  = \frac{1}{2}R_{\mu\nu}{}^{ab} M_{ab}
  + \tilde{R}_{\mu\nu}{}^{a} P_a
  + R_{\mu\nu}{}^{a} K_a
  + R_{\mu\nu} D ,
\end{equation}
the full component expressions, including the ordinary 4D curvature and torsion tensors, being given in~\cite{Roumelioti:2024lvn}.

Route II's dynamics follow from a curvature-squared, parity-preserving action
\begin{equation}
  S_{SO(2,4)}
  = a_{CG} \int d^4x \left[
    \operatorname{tr}\,\epsilon^{\mu\nu\rho\sigma} \, m \phi \, F_{\mu\nu}F_{\rho\sigma}
    + \lambda\left(\phi^2 - m^{-2}\mathbbm{1}_4\right)
  \right],
\end{equation}
built from a $\mathbf{15}$-valued scalar $\phi$, a dimensionful parameter $m$, a Lagrange multiplier $\lambda$, and a trace defined by $\mathrm{tr}\to \epsilon_{abcd}[\text{Generators}]^{abcd}$; being algebra-valued, $\phi$ itself expands as
\begin{equation}
  \phi = \phi^{ab} M_{ab} + \tilde{\phi}^{a} P_a + \phi^{a} K_a + \tilde{\phi} D .
\end{equation}

A convenient gauge choice~\cite{Li:1973mq} sets $\phi$ diagonal,
\[
  \phi = \mathrm{diag}(1,1,-1,-1),
\]
aligned entirely with the dilatation generator:
\begin{equation}
  \phi = \tilde{\phi}D \xrightarrow{\phi^2=m^{-2}\mathbb{I}_4} \phi = -2m^{-1}D .
\end{equation}
This gauge fixing is itself the SSB step, after which the action collapses to
\begin{equation}
  S = -2 a_{CG} \int d^4x\, \operatorname{tr}\,\epsilon^{\mu\nu\rho\sigma} F_{\mu\nu}F_{\rho\sigma}D ,
\end{equation}
and rescaling the gauge fields as \(e \to m e\), \(b \to m b\), and \(\tilde{a} \to m\tilde{a}\), then expanding \(F_{\mu\nu}\) through the conformal algebra, gives the Lorentz-invariant result~\cite{Roumelioti:2024lvn}
\begin{equation}
  \label{SO13action}
  S_{SO(1,3)}
  = \frac{a_{CG}}{4}\int d^4x\,
  \epsilon^{\mu\nu\rho\sigma}\epsilon_{abcd}\,
  R_{\mu\nu}{}^{ab}R_{\rho\sigma}{}^{cd}.
\end{equation}

Notice that $\tilde{a}_\mu$ has dropped out of the broken action entirely, so setting $\tilde{a}_\mu=0$ simplifies the $P$- and $K$-sector field strengths to
\begin{equation}
  \begin{aligned}
    \tilde{R}_{\mu\nu}{}^a
    &= m\, T_{\mu\nu}^{(0)a}(e) - 2m^2 \tilde{a}_{[\mu} e_{\nu]}{}^a
    \;\longrightarrow\;
    m\, T_{\mu\nu}^{(0)a}(e),\\[4pt]
    R_{\mu\nu}{}^a
    &= m\, T_{\mu\nu}^{(0)a}(b) + 2m^2 \tilde{a}_{[\mu} b_{\nu]}{}^a
    \;\longrightarrow\;
    m\, T_{\mu\nu}^{(0)a}(b),
  \end{aligned}
\end{equation}
with $T_{\mu\nu}^{(0)a}$ the usual Poincar\'e torsion tensor. Since neither combination enters the action, both may consistently be set to zero,
\[
  \tilde{R}_{\mu\nu}{}^{a} = 0,
  \qquad
  R_{\mu\nu}{}^{a} = 0,
\]
leaving a torsion-free theory. The dilatation curvature $R_{\mu\nu}$ is likewise absent from the action, and demanding $R_{\mu\nu}=0$ imposes the algebraic constraint
\begin{equation}
  \label{e-b-relation}
  e_\mu{}^a b_{\nu a} - e_\nu{}^a b_{\mu a} = 0
\end{equation}
between $e_\mu{}^a$ and $b_\mu{}^a$. Two natural ways of satisfying this relation are discussed next.

\paragraph{Case A: $\boxed{b_\mu{}^{a} = a\, e_\mu{}^{a}}\rightarrow$ Einstein Gravity}
First proposed in~\cite{Chamseddine:2002fd}, this ansatz turns~\eqref{SO13action} into, after some algebra,
\begin{equation}
  \begin{aligned}
    S_{SO(1,3)}
    = \frac{a_{CG}}{4}\int d^4x\, \epsilon^{\mu\nu\rho\sigma}\epsilon_{abcd}
    \Big[
      & R_{\mu\nu}^{(0)ab} R_{\rho\sigma}^{(0)cd}
      - 16 m^2 a\, R_{\mu\nu}^{(0)ab} e_\rho{}^c e_\sigma{}^d \\
      & + 64 m^4 a^2\, e_\mu{}^a e_\nu{}^b e_\rho{}^c e_\sigma{}^d
    \Big],
  \end{aligned}
\end{equation}
a sum of three pieces: a topological Gauss-Bonnet term with no effect on the field equations, the Einstein-Hilbert (Palatini) action in vierbein form, and a cosmological-constant contribution. Taking $a<0$ places the resulting theory on an AdS background.

\paragraph{Case B: $\boxed{b_\mu{}^{a} = -\frac{1}{4}\left(R_\mu{}^{a}-\frac{1}{6}R\, e_\mu{}^{a}\right)} \rightarrow$ Weyl Gravity}
This alternative choice~\cite{Kaku:1978nz, freedman_vanproeyen_2012}, inserted into~\eqref{SO13action}, instead produces an action built from a pair of Weyl conformal tensors $C_{\mu\nu}{}^{ab}$:
\begin{equation}
  \label{Weyl1}
  S = \frac{a_{CG}}{4} \int d^4x\, \epsilon^{\mu\nu\rho\sigma}\epsilon_{abcd}\, C_{\mu\nu}{}^{ab} C_{\rho\sigma}{}^{cd}\ ,
\end{equation}
or, equivalently, the familiar four-dimensional scale-invariant Weyl action
\begin{equation}
  \label{Weyl2}
  S_W =2a_{CG}\int \mathrm{d}^4 x\left(R_{\mu \nu} R^{\nu \mu}-\frac{1}{3} R^2\right).
\end{equation}
Scale invariance forbids a cosmological-constant term here, but otherwise WG shares CG's appeal as a candidate high-energy completion of gravity (see, e.g.,~\cite{Maldacena:2011mk, mannheim, Anastasiou:2016jix, ghilencea2023, Hell:2023rbf, Condeescu:2023izl}).

One can ask the converse question - whether EG can be recovered from WG by a further SSB step - and the answer is yes: a vector-representation ($\mathbf{6}$) scalar of $SU(4)\sim SO(2,4)$ again does the job, breaking WG down to the Einstein-Hilbert action~\cite{Patellis:2025qbl, Patellis:2025syq}.

It is also worth stressing~\cite{Patellis:2025qbl} that the on-shell vanishing of both torsions $\tilde{R}_{\mu\nu}{}^{a}$ and $R_{\mu\nu}{}^{a}$, together with the curvature $F_{\mu \nu}$, is precisely what guarantees that diffeomorphisms and gauge transformations coincide in this framework.

Equivalently, this same endpoint follows from the two-vector-scalar mechanism of Route~I above.

\section{Gauge-Theoretic Formulation of Noncommutative (Fuzzy) Gravity}

\subsection{The Noncommutative Background Space}

Formulating fuzzy gravity as a gauge theory first requires fixing the noncommutative background it lives on. Building on Snyder's and Yang's original construction~\cite{Snyder:1946qz, yang1947} and its later extensions~\cite{Heckman_2015, Manolakos_paper1, Manolakos_paper2, Manolakos:2022universe, roumelioti2407}, spacetime is taken to be a noncommutative manifold whose coordinate operators descend from the Lie algebra of $SO(1,5)$.

Writing $J_{mn}$, with indices $m,n,r,s=0,\dots,5$, for the generators of $SO(1,5)$ (metric $\eta_{mn}=\mathrm{diag}(-1,1,1,1,1,1)$), the algebra reads
\begin{equation}
  \left[J_{mn}, J_{rs}\right]
  = i \left(
    \eta_{mr} J_{ns} + \eta_{ns} J_{mr}
    - \eta_{nr} J_{ms} - \eta_{ms} J_{nr}
  \right).
\end{equation}

Recovering a four-dimensional geometric interpretation requires descending through the subgroup chain
\[
  SO(1,5) \supset SO(1,4) \supset SO(1,3),
\]
under which the algebra decomposes as
\begin{equation}
  \begin{gathered}
    \left[J_{ij},J_{kl}\right]
    = i \left(
      \eta_{ik}J_{jl} + \eta_{jl}J_{ik}
      - \eta_{jk}J_{il} - \eta_{il}J_{jk}
    \right), \qquad
    \left[J_{ij}, J_{k5}\right]
    = i \left(\eta_{ik}J_{j5} - \eta_{jk}J_{i5}\right), \\[4pt]
    \left[J_{i5}, J_{j5}\right] = i J_{ij}, \qquad
    \left[J_{ij}, J_{k4}\right]
    = i \left(\eta_{ik}J_{j4} - \eta_{jk}J_{i4}\right), \qquad
    \left[J_{i4}, J_{j4}\right] = i J_{ij}, \\[4pt]
    \left[J_{i4}, J_{j5}\right] = i \eta_{ij} J_{45}, \qquad
    \left[J_{ij}, J_{45}\right] = 0, \qquad
    \left[J_{i4}, J_{45}\right] = -i J_{i5}, \qquad
    \left[J_{i5}, J_{45}\right] = i J_{i4}.
  \end{gathered}
\end{equation}

A physical reading of these generators emerges by identifying them with the noncommutativity tensor, spacetime coordinates and momenta via
\begin{equation}
  \Theta_{ij} = \hbar J_{ij}, \qquad
  X_i = \lambda J_{i5}, \qquad
  P_i = \frac{\hbar}{\lambda} J_{i4}, \qquad
  h = J_{45},
\end{equation}
where $\lambda$ fixes a length scale. Feeding this dictionary back into the algebra produces the operator relations
\begin{equation}
  \begin{gathered}
    \relax[\Theta_{ij}, \Theta_{kl}]
    = i\hbar \left(
      \eta_{ik}\Theta_{jl} + \eta_{jl}\Theta_{ik}
      - \eta_{jk}\Theta_{il} - \eta_{il}\Theta_{jk}
    \right), \\[4pt]
    [\Theta_{ij}, X_k] = i\hbar (\eta_{ik} X_j - \eta_{jk} X_i),
    \qquad
    [\Theta_{ij}, P_k] = i\hbar (\eta_{ik} P_j - \eta_{jk} P_i), \\[4pt]
    [X_i, X_j] = \frac{i\lambda^2}{\hbar} \Theta_{ij}, \qquad
    [P_i, P_j] = \frac{i\hbar}{\lambda^2} \Theta_{ij}, \qquad
    [X_i, P_j] = i \hbar \eta_{ij} h, \\[4pt]
    [\Theta_{ij}, h] = 0, \qquad
    [X_i, h] = \frac{i\lambda^2}{\hbar} P_i, \qquad
    [P_i, h] = -\frac{i\hbar}{\lambda^2} X_i.
  \end{gathered}
\end{equation}

Two physical features stand out in this algebra: the spacetime coordinates fail to commute,
\[
  [X_i, X_j] \propto \Theta_{ij},
\]
with momenta obeying an analogous relation - signalling a discrete, quantised structure on both sides - while the mixed commutator $[X_i,P_j]$ reduces to a generalised Heisenberg relation.

The Snyder-Yang algebra is therefore a natural, self-consistent stage on which to build a gauge theory of fuzzy gravity.

\subsection{Gauge Theory of Fuzzy Gravity}

Gauging this background starts with choosing the gauge group. The obvious candidate is $SO(1,4)$, the isometry group of $dS_4$ and hence the symmetry of the corresponding commutative theory; but noncommutative gauge theories unavoidably generate anticommutators of the gauge generators alongside commutators, and $SO(1,4)$'s anticommutators do not close back within the same algebra. A representation closed under both operations is therefore required.

The resolution~\cite{Manolakos_paper1, Manolakos_paper2} is to fix a suitable representation and enlarge $SO(1,4)$ to $SO(2,4)\times U(1)$ - the smallest extension whose generators close under both brackets - which then serves as the gauge group for fuzzy gravity on the covariant Snyder-Yang background.

\vspace{0.3cm}

The covariant coordinate is introduced as
\begin{equation}\label{CovariantCoordinate}
  \mathcal{X}_\mu = X_\mu \otimes \mathbbm{1}_4 + A_\mu(X),
\end{equation}
with $A_\mu$ the noncommutative gauge connection; expanded on the generators of $SO(2,4)\times U(1)$,
\begin{equation}\label{GaugeConnectionFuzzy}
  A_\mu
  = a_\mu \otimes \mathbbm{1}_4
  + \omega_\mu{}^{ab}\otimes M_{ab}
  + e_\mu{}^a \otimes P_a
  + b_\mu{}^a \otimes K_a
  + \tilde{a}_\mu \otimes D,
\end{equation}
it turns~\eqref{CovariantCoordinate} into
\begin{equation}
  \mathcal{X}_\mu
  = (X_\mu + a_\mu)\otimes \mathbbm{1}_4
  + \omega_\mu{}^{ab}\otimes M_{ab}
  + e_\mu{}^a \otimes P_a
  + b_\mu{}^a \otimes K_a
  + \tilde{a}_\mu \otimes D.
\end{equation}

Its field strength is defined by~\cite{Madore_1992, Manolakos_paper1}
\begin{equation}
  \hat{F}_{\mu\nu} \equiv [\mathcal{X}_\mu, \mathcal{X}_\nu] - \kappa^2 \hat{\Theta}_{\mu\nu},
\end{equation}
where
\[
  \hat{\Theta}_{\mu\nu} \equiv \Theta_{\mu\nu} + \mathcal{B}_{\mu\nu}
\]
absorbs a two-form $\mathcal{B}_{\mu\nu}$ needed to keep $\Theta_{\mu\nu}$ covariant; being algebra-valued, $\hat{F}_{\mu\nu}$ decomposes as
\begin{equation}
  \hat{F}_{\mu\nu}= R_{\mu\nu}\otimes \mathbbm{1}_4 + \frac{1}{2} R_{\mu\nu}{}^{ab}\otimes M_{ab} + \tilde{R}_{\mu\nu}{}^a \otimes P_a + R_{\mu\nu}{}^a \otimes K_a + \tilde{R}_{\mu\nu}\otimes D.
\end{equation}

As in the conformal case, a physically sensible theory requires an SSB step: a scalar $\Phi(X)$ in the second-rank antisymmetric representation of $SO(2,4)$, carrying in addition a $U(1)$ charge, is introduced to break the extended symmetry completely; a suitable gauge choice for $\Phi$ leaves $SO(1,3)$ as the unbroken local symmetry~\cite{Manolakos_paper1, Manolakos_paper2, Roumelioti:2024lvn}.

This yields the fuzzy-gravity action
\begin{equation}
  \mathcal{S} = \operatorname{Trtr} \Big[ \lambda\, \Phi(X)\, \varepsilon^{\mu\nu\rho\sigma} \hat{F}_{\mu\nu}\hat{F}_{\rho\sigma} + \eta\Big( \Phi(X)^2 - \lambda^{-2}\mathbbm{1}_N\otimes\mathbbm{1}_4 \Big) \Big],
\end{equation}
with the outer trace running over the coordinate matrices and the inner one over the gauge-group generators, $\eta$ a Lagrange multiplier and $\lambda$ a dimensionful parameter; $SO(1,3)$ survives as the unbroken symmetry. Taking the commutative limit reproduces the Palatini action~\cite{Manolakos_paper2}, so ordinary Einstein gravity with a cosmological constant is recovered, exactly as one would expect.

\section{SO(2,16) Unification of Gravities and Internal Interactions}
\label{sec4}

The construction of Sec.~\ref{sec2} unifies gravity with an $SO(10)$ GUT by enlarging the tangent-space symmetry all the way up to $SO(2,16)$~\cite{Roumelioti:2024lvn}. Two requirements fix this choice: the unification group must admit an SSB chain reaching both $SO(2,4)$ and $SO(10)$, and, for the resulting fermion spectrum to be chiral, it must take the form $SO(4n+2)$. Scanning groups of this form for the smallest one compatible with both conditions we are lead to $SO(2,16)$. This is simply a higher-dimensional instance of the theme running through this whole review~\cite{Weinberg:1984ke, roumelioti2407, Percacci:1984ai, Percacci_1991, Nesti_2008, Nesti_2010, Krasnov:2017epi, Chamseddine2010, Chamseddine2016, noncomtomos, Konitopoulos:2023wst}: the tangent group need not have the same dimension as the manifold it acts on.

Working in Euclidean signature for convenience, the starting point is $SO(18)\sim SO(2,16)$ with fermions already restricted to the Weyl condition, in the spinor representation $\mathbf{256}$. A first SSB step takes $SO(18)$ down to its maximal subgroup $SO(6)\times SO(12)$, with the relevant representations branching as
\begin{equation}\label{so18-unification}
  \begin{aligned}
    SO(18) & \supset SO(6) \times SO(12) \\
    \mathbf{256} & = (\mathbf{4}, \overline{\mathbf{32}}) + (\overline{\mathbf{4}}, \mathbf{32})
    && \text{(spinor)} \\
    {\mathbf{153}} & =(\mathbf{15}, \mathbf{1}) + (\mathbf{6}, \mathbf{12}) + (\mathbf{1}, \mathbf{66}) & & \text {(adjoint)} \\
    \mathbf{170} & = (\mathbf{1}, \mathbf{1}) + (\mathbf{6}, \mathbf{12}) + (\mathbf{20}', \mathbf{1}) + (\mathbf{1}, \mathbf{77})
    && \text{(2nd rank symmetric)}\, .
  \end{aligned}
\end{equation}
This particular breaking is triggered by a vev of a $\mathbf{170}$-scalar pointing along its $\langle(\mathbf{1},\mathbf{1})\rangle$ singlet direction, while the fermions remain in the $\mathbf{256}$ spinor throughout this step.

The $SO(12)$ factor can subsequently be reduced to $SO(10)$ together with either a gauged or a residual global $U(1)$, and which outcome is realised depends on where the symmetry-breaking scalar sits: a vev for the $\langle(\mathbf{1})(0)\rangle$ singlet inside the $\mathbf{66}$ (part of the adjoint $\mathbf{153}$ of $SO(18)$) leaves the $U(1)$ gauged, whereas a vev for the $\langle(\mathbf{1})(4)\rangle$ singlet inside the $\mathbf{77}$ (part of the symmetric $\mathbf{170}$) breaks it to a global symmetry instead. The two branchings responsible for this are
\begin{equation}
  \begin{aligned}
    SO(12) & \supset SO(10) \times \left[U(1)\right]\\
    \mathbf{66} & =(\mathbf{1})(0)+( \mathbf{10})(2)+( \mathbf{10})(-2)+( \mathbf{45})(0) \\
    \mathbf{77} & =(\mathbf{1})(4)+( \mathbf{1})(0)+( \mathbf{1})(-4)+( \mathbf{10})(2)+( \mathbf{10})(-2)+( \mathbf{54})(0)\, ,
  \end{aligned}
\end{equation}
with $\left[U(1)\right]$ denoting that either outcome is possible depending on the vacuum chosen.

The complementary $SO(6)$ factor is reduced to the Lorentz group by exactly the two-step pattern already used for Route~I of Sec.~\ref{sec2}~\cite{Slansky:1981yr}: identifying $SO(6)\sim SU(4)$, eq.~\eqref{SO6toSO5} first drives $SU(4)\to SO(5)$ via a singlet vev in the $\mathbf{6}$, after which eq.~\eqref{SO5toSU2SU2} completes $SO(5)\to SU(2)\times SU(2)\sim SO(1,3)$. Under this chain the fundamental $\mathbf{4}$ splits precisely into the two Weyl-spinor representations that together assemble a four-dimensional Dirac fermion.

A second, equally valid route to the same endpoint mirrors the alternative used for CG: rather than the vector representation, one uses a scalar in the adjoint $\mathbf{15}$ of $SU(4)$ (itself sitting inside the adjoint $\mathbf{153}$ of $SO(18)$), giving the branching
\begin{equation}
  \begin{aligned}
    SU(4) \supset & SU(2) \times SU(2) \times U(1)\\
    \textbf{4} = & (\textbf{2},\textbf{1})(1)+(\textbf{1},\textbf{2})(-1)\\
    \textbf{15}  = & (\textbf{1},\textbf{1})(0)+(\textbf{2},\textbf{2})(2)+(\textbf{2},\textbf{2})(-2)\\
    &+(\textbf{3},\textbf{1})(0)+(\textbf{1},\textbf{3})(0)\, ,
  \end{aligned}
\end{equation}
A vev along $(\mathbf{1},\mathbf{1})$ in the adjoint $\mathbf{15}$ reduces $SU(4)$ to $SU(2)\times SU(2)\times U(1)$~\cite{Li:1973mq}, and the spurious $U(1)$ gauge boson is disposed of exactly as before, leaving $SU(2)\times SU(2)$; once again the $\mathbf{4}$ organises itself into the two Weyl spinors expected of a Dirac fermion in four dimensions. Since $SO(2,16)\sim SO(18)$ and $SO(2,4)\sim SO(6)\sim SU(4)$ are non-compact/compact realisations of the same algebras, both routes above apply equally well whichever signature is used.

Because $SO(18)$ is a $4n+2$ group, chirality survives dimensional reduction: starting from a single Weyl spinor in the $\mathbf{256}$, tracking it through both SSB chains above gives, after all breakings,
\begin{equation}
  \begin{gathered}
    SU(2)\times SU(2) \times SO(10) \times [U(1)]\\
    \{(\textbf{2},\textbf{1})+(\textbf{1},\textbf{2})\}\{\textbf{16}(-1)+\overline{\textbf{16}}(1)\}+\{(\textbf{2},\textbf{1})+(\textbf{1},\textbf{2})\}\{\overline{\textbf{16}}(1)+\textbf{16}(-1)\}\\
    =2\times\textbf{16}_L(-1)+2\times\overline{\textbf{16}}_L(1)+2\times\textbf{16}_R(-1)+2\times\overline{\textbf{16}}_R(1)\, ,
  \end{gathered}
\end{equation}
and identifying $\overline{\textbf{16}}_R(1)=\textbf{16}_L(-1)$, $\overline{\textbf{16}}_L(1)=\textbf{16}_R(-1)$, keeping only the $\gamma^5=-1$ eigenspace collapses this to
\begin{equation}
  4\times \textbf{16}_L(-1)\, .
\end{equation}
Four copies of the $\mathbf{16}_L$ of $SO(10)$ is precisely the fermion content of four Standard-Model-like families, so the group-theoretic structure of the $SO(2,16)$ construction fixes the family number without any additional input. This multiplicity could in principle be halved by a Majorana condition, but that option is unavailable here: for an $SO(t,s)$ group (with $t$ the number of time directions and $s$ the number of space ones), compatibility of Majorana and Weyl conditions requires $s-t=0\mod 8$, which $SO(2,16)$ does not satisfy. Separating the four families into distinct flavours remains an open problem, taken up again in Sec.~\ref{sec6}.

Unifying fuzzy gravity (FG) with internal interactions imposes two extra constraints on where the fermions can live, as emphasized in~\cite{roumelioti2407}: they must remain chiral (so as not to pick up Planck-scale masses) and must be organised in matrix representations compatible with the FG matrix-model construction. Ref.~\cite{Chatzistavrakidis:2010xi, Chatzistavrakidis:2011toc} showed that both requirements can be met simultaneously by placing fermions in bifundamental, rather than fundamental or tensor, representations of the relevant product gauge group — a strategy that works equally for $SU(N)$ and $SO(N)$ factors and has recurred in several other contexts~\cite{Ibanez:1998xn, Ma:2004mi, Leontaris:2005ax, Irges:2011de, Manolakos:2020cco, Patellis:2024dfl}. Here this is implemented starting from $SO(6)\times SO(12)$ with fermions in $(\mathbf{4}, \overline{\mathbf{32}})+(\overline{\mathbf{4}}, \mathbf{32})$, which satisfies both conditions at once. Since the FG gauge group is $SO(2,4)\times U(1)\sim SO(6)\times U(1)$, the resulting low-energy structure parallels the CG case discussed above.

\section{Conclusions and Discussion on Future Prospects}
\label{sec6}

Across Secs.~\ref{sec2}-\ref{sec4} we have reviewed how gauging an enlarged tangent-space symmetry~\cite{Roumelioti:2024lvn} allows gravity and internal interactions to be unified within a single four-dimensional gauge theory, the guiding idea throughout being that the tangent group's dimension need not match that of the manifold it acts on. Starting from $SO(2,4)$, conformal gravity emerges as a gauge theory whose SSB reduces it, depending on the vacuum chosen, to either Einstein or Weyl gravity as a low-energy limit. Enlarging the tangent group further to $SO(2,16)$ then incorporates internal interactions through an $SO(10)$ GUT, with fermions constrained by the Weyl condition, while the parallel fuzzy-gravity construction~\cite{roumelioti2407} reaches the same goal starting instead from $SO(2,4)\times SO(12)$ with fermions in $(\mathbf{4}, \overline{\mathbf{32}})+(\overline{\mathbf{4}}, \mathbf{32})$.

The phenomenology of this $SO(2,16)$ scheme has already been examined at one loop~\cite{Patellis-Z-24}: four candidate symmetry-breaking chains from $SO(10)$ to the Standard Model were studied, giving estimates for the breaking scales between the Planck and Einstein-gravity regimes, together with the gravitational-wave signal such chains would leave via cosmic strings~\cite{Patellis:2025qbl}.

Building on this picture, several directions remain open. An obvious one is to construct a minimal unification scheme that unifies Einstein Gravity with internal interactions. Given also the difficulty described earlier in Sec.~\ref{sec4} of obtaining eventually a degeneracy of four fermion families, we suggest a unified scheme in which the Majorana condition could be applied in addition to the Weyl one, which would divide the multiplicity of the number of families by two. The condition for having Weyl-Majorana spinors in an $SO(t,s)$ gauge theory is $(s-t)=0 \mod 8$. Combining the Weyl-Majorana condition with the requirements that (a) the unification group should admit spontaneous symmetry breaking (SSB) chains leading to both $SO(1,3)$ and $SO(10)$, and (b) the theory should allow a chiral structure, realised for groups of the form $4n+2$, as already stated in Sec.~\ref{sec4}, we are led to consider the gauge theory based on $SO(1,17)$ as the minimal gauge unification group. We then expect to obtain two fermion families after all SSB. This degeneracy is easily lifted by using scalars in different representations that provide masses to the fermions through their common Yukawa coupling arising from $SO(10)$ unification. The prediction of this unified scheme is therefore an even number of families, with a minimum of four~\cite{Stefas-Zoupanos-inprep}.

Going further, in studying the unification of all interactions based on $SO(1,17)$, our plan is to study the various scales appearing in this unification scheme using renormalization-group analysis, in analogy to that carried out in~\cite{Patellis-Z-24}. This will in turn require a fresh examination of the $SO(10)$ spontaneous-symmetry-breaking patterns towards the SM against proton-lifetime experimental bounds, as well as their observational potential through the gravitational-wave signal from cosmic-string production, as discussed in~\cite{Patellis:2025qbl}.

Next, the unification schemes of conformal and fuzzy gravities with internal interactions that have been constructed require further examination of their cosmological predictions. We recall that in these schemes, the various stages of SSB leave behind several heavy particles whose interaction with ordinary matter is Planck-suppressed. Depending on the SSB scales, the resulting heavy particles can become cosmologically stable, making them excellent Dark Matter candidates. We plan to examine this scenario in detail. Realising it requires introducing gauge-invariant terms for the scalars responsible for the SSB in the action of the models, i.e. promoting the so-far auxiliary scalars to dynamical ones. This can be done without introducing ghosts in a dS background, by adjusting the AdS background arising from the gravity side against the positive contribution from the vevs of the various scalar fields. Therefore, from our point of view, a very interesting suggestion is to regard the second vielbein - the special conformal gauge field - in our models as a dark matter field. In particular it will be shown that this field is expected to be stable (or sufficiently long-lived) since its interactions with the visible sector are expected to be highly suppressed after symmetry breaking~\cite{Nandi-Stefas-Zoupanos-inprep}.

Finally, let us discuss the ghost problem in the presented framework and prospects for its resolution. The inclusion of higher-curvature terms in the action introduces higher derivatives of the graviton field, which in turn give rise to a massive spin-2 Ostrogradsky ghost. As first demonstrated in the seminal work of Stelle~\cite{stelle.77,Stelle:1977ry}, a purely quadratic curvature action in four dimensions is power-counting renormalizable; however, it is non-unitary due to the propagation of a spin-2 ghost degree of freedom. This tension between renormalizability and unitarity poses a major obstacle, suggesting that a consistent perturbative theory of quantum gravity may be unattainable. The construction of gravity theories discussed above obviously belongs to this category, as does the celebrated Starobinsky inflationary model~\cite{Starobinsky:1980te}, where a Ricci-scalar-squared term is added to the Einstein-Hilbert action.

Several solutions have been proposed (see e.g.~\cite{Lambiase:2025qhb}), but none of these have been sufficiently convincing for the theory to be accepted as a consistent alternative to GR. Among these, of particular interest is the requirement of adding specific boundary conditions~\cite{Maldacena:2011mk, Hell:2023rbf}. However, the proposal is purely classical, or tree-level, and it would be extremely interesting if one could use it to construct a full quantum theory based on conformal gravity. This is certainly one direction of further research.

In the presence of ghost, if its contribution becomes dominant, the theory lies outside the domain of validity of the effective field theory (EFT), leading to a loss of theoretical control. Predictivity can nevertheless be maintained if the ghost mass, when present, is much larger than the relevant energy scale.

On the other hand, interactions for massive spin-2 fields have long been thought to inevitably give rise to ghost instabilities due to the Boulware-Deser ghost~\cite{Boulware:1972yc}. However, a ghost-free nonlinear potential for the massive spin-2 field was found by de Rham et al.~\cite{deRham:2010kj,deRham:2010tw}. This led to the recent breakthrough in the physics of gravitation with the construction of ghost-free bimetric theory (see~\cite{SchmidtMay:2015vnx} for a review). This theory contains, in addition to the usual massless graviton, a second propagating spin-2 particle with non-zero mass.

Theories with extended symmetries are known to generically exhibit improved quantum behaviour. An example for the spin-2 case is the action for conformal gravity with its Weyl symmetry. Unfortunately, as already noted, this theory suffers from a ghost, but bimetric theory contains a ghost-free model that appears closely related to the conformal and Weyl-invariant actions considered here. Therefore, the two well-known sources of ghosts may be related within the conformal gravity framework studied in this work.

Understanding further the properties of ghosts in our constructions could provide new and important insights into the nature of quantum gravity. We plan an exhaustive study within our construction to resolve this long-standing problem.

\section*{Acknowledgements}
We are grateful to Vasilis Letsios, Pantelis Manousselis, and Partha Nandi for numerous illuminating discussions during the development of the work presented in this paper. GZ would like to thank the organizers of SAGS2025 for the excellent hospitality that was offered to him.

\bibliographystyle{unsrt}
\bibliography{bibliography}

\end{document}